\documentstyle[epsfig]{mn}
\input epsf.tex
\title{Cosmological GRBs: Internal vs. External Shocks}
\author[Sari \& Piran]{Re'em Sari \& Tsvi Piran\\
	Racah Institute for Physics, The Hebrew University,
        Jerusalem 91904, Israel}
\def\etal{{\it et. al., }}

\input mn_math.sty
\begin{document}

\maketitle
\begin{abstract}
Internal (IS) or external (ES) shocks in relativistic expanding shells
are currently the best known mechanism for producing GRBs.  We
calculate the hydrodynamic conditions and the cooling processes in
internal shocks, that occur when one layer overtakes another.  We
compare these conditions with the conditions in external shocks that
occur when the shell encounters the ISM. We find that the complicated
temporal structure observed in GRBs could be easily explained in IS
scenario. Unlike ES, the observed temporal structure simply reflects
the temporal behavior of the internal engine that produces the
shell. We also find that opacity and in particular the opacity for
pair production poses strong constraints on the parameter space and
consequently bursts with very narrow peaks are expected to be
optically thick and would not contain GeV photons. Finally we find
that as in ES synchrotron is the most likely radiation process.

\end{abstract}

\centerline{\it Subject heading: gamma-rays: bursts-hydrodynamics-relativity}

\section{Introduction}

A cosmological $\gamma$-ray burst (GRB) occurs, most likely, when the
kinetic energy of a relativistic shell turns into internal energy and
cools by non-thermal radiation process. The kinetic energy of the
shell can be transformed into internal energy by decelerating onto the
ISM (M\'esz\'aros \& Rees 1992), or by internal shocks in the shell
(Rees \& M\'esz\'aros  1994; Narayan, Paczy\'nski \& Piran, 1992).

Sari \& Piran (1995, denoted hereafter SP) have discussed the
hydrodynamics of the deceleration of the relativistic shell on the ISM
and internal shocks.  If the shell is thin and its Lorentz factor is
low, internal shocks form before a considerable deceleration on the
ISM takes place. Provided that the Lorentz factor of the shell is not
uniform, with fluctuations of the order of a few, these internal shocks
can extract a considerable fraction of the shell's  kinetic energy
and produce the observed $\gamma$-rays.

The internal shock (IS) scenario has several advantages over the
external shock (ES) one.  M\'esaz\'aros and Rees (1994) have pointed
out that IS requires, quite generally, a lower Lorentz factor. This
allows slightly larger amounts of baryonic load in the fireball and it
eases, somewhat, the constraints on the source. We demonstrate in
sections II and III that the hydrodynamics of IS also offers a natural
explanation to the multiple time scales observed in the temporal
profiles of the bursts. Such an explanation is missing in the ES
scenario.

The observed spectrum indicates that the source must be optically
thin. Two processes are relevant for determining the optical depth:
Compton scattering on the shell's electrons and pair production
between the $\gamma$-ray photons. The relativistic motion of the shell
releases the constraint that the optical depth poses to cosmological
GRB sources. In fact this is the main reason to believe that highly
relativistic motion is an essential ingredient of any cosmological GRB
model (Goodman, 1986, Paczy\'nski 1986; Krolik and Pier, 1989; Piran,
1996). For the ES scenario in which the kinetic energy is deposited at
$\sim 10^{15}-10^{16}$cm the optical thinness condition leads to a
modest limitation on the Lorentz factor $\gamma \ge 100$ (Fenimore
Epstein \& Ho, 1993; Woods \& Loeb 1995; Piran 1996) which is not a
serious restriction. For ES a stronger limit on $\gamma$ arises from
the temporal structure.  For internal shocks, in which the energy is
deposited around $\sim 10^{12}-10^{14}$cm this poses an additional
constraint which we examine in section IV. In the IS scenario the
minimal Lorentz factor is determined by optical thinness condition.
We show (in section IV) that the optical depth to pair production
poses, usually a stronger constraint than the optical depth to Compton
scattering which M\'esaz\'aros and Rees (1994) used.

Cooling must be rapid enough to allow for the shortest time scale.  In
a recent paper, Sari, Narayan and Piran (1996 denoted hereafter SNP)
have discussed cooling in the ES scenario.  In sections V and VI we
generalize this discussion to internal shocks, and we discuss the
possible observational differences between cooling in the ES and the
IS scenarios.

\section{Temporal Structure}

Many bursts display a complicated temporal structure with numerous
peaks (see e.g. Fishman and Meegan, 1995).  We denote by $t_{dur}$
the total duration of the burst and by $\delta t$ the typical duration
of the individual peaks.  Few bursts are smooth with $\delta t \approx
t_{dur}$.  Most bursts are highly variable with $\delta t \ll \Delta
T$. Thus any GRB model must be capable of explaining both time scales,
the overall duration, $t_{dur}$, and the duration of the individual
peaks, $\delta t$. We review here, briefly, several relativistic
effects that determine the temporal structure of the bursts (see
Fenimore, Madras and Nayakshin, 1996  for a detailed discussion).

Consider a shell with a width $\Delta$ (in the observer frame) and a
Lorentz factor $\gamma$ that converts its kinetic energy to internal
energy and cools between $R_e$ and $2R_e$ (see Fig. 1).  A photon that
is emitted at the outer edge of the shell will reach the observer
$\Delta /c$ before a  photon that is emitted at the same time (in
the observer's rest frame) from  the inner edge of the shell. Therefore:
\begin{equation}
t_{dur} \ge \Delta /c \ .
\end{equation}

We denote by $\gamma_2$ the Lorentz factor of the radiating matter
($\gamma_2$ is always smaller than $\gamma$).  A photon that is
emitted by a given electron at $R_e$ will arrive $R_e/2 \gamma_2^2 c$
before a photon emitted by the same electron at $2R_e$ (the difference
arises from the the different time of flight of the photon and the
electron from $R_e$ to $2R_e$) (see Fig. 1).  Finally the angular
width of the observed emitting surface is $1/\gamma_2$ due to
relativistic beaming. A photon that it emitted radially outwards will
reach the observer $R_e/\gamma_2^2 c$ before a photon that is emitted
by matter that moves at an angle $1/\gamma_2$ from the line of sight
(Katz, 1994) (see Fig. 1). The last two limits give rise to the same
condition, which limits the duration of peaks.  If the emission is
continuous from $R_e$ to $2R_e$ or if it spreads over a region with an
angular width larger than $R_e/\gamma_2$ then:
\begin{equation}
\delta  t \ge R_e /  \gamma_2^2  c \ .
\end{equation}

\begin{figure}
\psfig{file=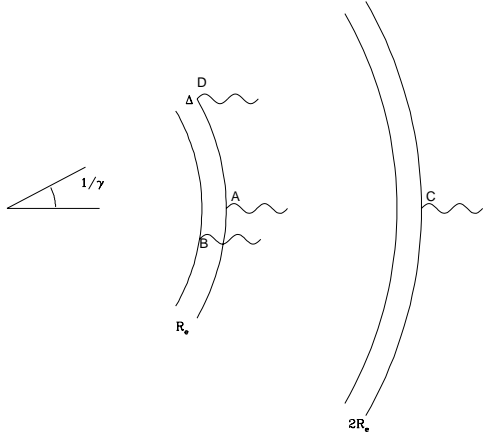,width=220pt}
\caption{A photon emitted at (A) the outer side of a shell with thickness
$\Delta$ will reach the oberve $\Delta /c$ before a photon emitted at
the same time (in the obesrver's frame) from  the inner side of the shell
(B). A photon emitted at (A) will reach the observer $R_e/2\gamma^2 c$ 
before a photon emitted by the same electron at $2r_e$ (C). A photon
emitted at (A) will reach the observer   $R_e/\gamma^2 c$ before
a photon emitted by matter that moves radially outwards at an angle
$1/\gamma$.}
\label{fig1}
\end{figure}

Within the ES scenario there are two possible regimes for deceleration
of the shell on the ISM according to the value of the parameter $\xi$
(SP):
\begin{equation}
\xi\equiv (l/\Delta)^{1/2}\gamma^{-4/3} \ , 
\end{equation}
where $l \equiv (E/n_{ism} m_{p}c^2)^{1/3}$ is the Sedov length
typically $\sim 10^{18}$cm. If $\xi>1$ the shock penetrating the shell
is only mildly relativistic, the shell decelerates gradually, and
$\gamma_2 \cong \gamma$. We call this case the ``Newtonian'' scenario.
If $\xi < 1$ the shell is decelerated by one ultra relativistic shock,
which reduces its Lorenz factor from $\gamma$ directly to $\gamma_2=
\gamma \xi ^{3/4}$. This is the relativistic scenario.

In either case the external shock satisfies: $\Delta \le
R_e/\gamma_2^2$.  Thus, if the energy conversion in the shell and the
subsequent emission is continuous or if the emission spreads over an
angular size $R_e/\gamma_2$ then $t_{dur} \approx \delta t \approx
dt$.  This corresponds to a smooth temporal profile with a single
peak.  Such profiles are observed only in a small minority of the
bursts.  This conclusion is quite severe. It suggests that angular
spreading will smooth all bursts and ES are not capable of producing
bursts with complicated time profiles.

A way out of the angular spreading problem, within the external shock
scenario, is if the ISM is made out of small dense clouds with a large
filling factor.  There is an enhanced emission when the shell reaches
ISM regions with higher density. This requires large fluctuations in
the ISM on the scale of $\approx 10^{12}$cm (for peaks with 
$\delta t \approx 0.3$ sec and $\gamma=100$).  Shaviv and Dar (1995,1996)
suggest and explore in details the temporal structure for the scenario
in which the shell interacts with a dense population of stars (in a
globular cluster or a galactic center) which provide the desired high
filling factor.  Recently, Li \& Fenimore (1996) have found that the
distribution of peaks in long GRBs does not correspond to a random
distribution.  More specifically, the intervals between peaks has a
log-normal distribution.  Such a distribution is incompatible with a
random distribution of objects that the shell encounters. We have
already seen that the inner source cannot produce variability on such
scale (because of the angular spreading).  Hence, this feature, if
confirmed, would poses a possible objection to the ES model.
\footnote{An alternative solution to the angular spreading problem
would be if the source produces an irregular shell composed of many
small blobs.  Internal pressure will cause any small scale angular
structure in the shell to spread radially and tangentially to size
$R/\gamma$ (Narayan and Piran, 1994, Piran 1994). However, recall 
that $\gamma_2 \ll \gamma$ in the relativistic ES scenario. If the
shocked material cools  before it expands from $R_e/\gamma$ to
$R_e/\gamma_2$ it will  produce a short peak  on a scale
$t_{dur}/c) \gamma_2/\gamma$ (Sari and Piran, 1996). Temporal
correlation in this substructure could be produced naturally by the
internal engine}

The essence of the problem in the external shock scenario was the
inequality: $\Delta \le R_e/\gamma_2^2$. We will see shortly that
within the IS scenario we have, quite generally, $\Delta >
R_e/\gamma_2^2$ and $\gamma_2 \approx \gamma$.  Consequently in this
case $\Delta$ determines the overall duration, $t_{dur}$, while the
angular spreading as well as the radial size of the shell's
fluctuations determines the duration of individual peaks, $\delta
t$. This, in turn, yields some interesting information about the inner
source that drives GRBs. It requires that the source produces a shell
of relativistic matter of width $\Delta$ in which fluctuations of
order unity in the Lorentz factor should take place on a scale $\delta
\ll \Delta$.

\section{Hydrodynamic of Internal Shocks}

Internal shocks occur when the relativistic ejecta from the inner
source that drives a GRB is not moving uniformly. If some inner layer
moves faster than an outer one it will take over.  When the two layers
collide two shocks appear - a forward shock propagating into the outer
shell and a reverse shock propagating into the inner one. The forward
and the reverse shocks have similar conditions behind them since it is
most likely that both layers had (prior to the collision) similar
conditions.  This simplifies the discussion of internal shocks in
which the two emitting regions have similar physical condition.  In
the case of external shocks we had to deal with each shock separately
(SNP).

We begin with a short discussion of the hydrodynamics of internal
shocks in relativistic shells. Imagine a spherically expanding
relativistic shell of thickness $\Delta$ in the observer frame moving
with a high Lorentz factor $\gamma$. Assume further that the Lorentz
factor varies  by order of a few throughout the shell.  Faster
shells overtake  slower ones. An inner shell will overtake an outer
one at a radius, $R_s$:
\begin{equation}
R_s \sim \delta \gamma^2 \ ,
\label{rs}
\end{equation}
where $\delta$ is the spatial scale of fluctuations in $\gamma$
within the shell, and we have ignored factors of order unity.  A
significant fraction of the kinetic energy is converted to internal
energy at this radius.

The observed time over which internal shocks transform the relative 
kinetic energy of the two colliding layers (whose typical size is
$\sim \delta$) is of the order of
\begin{equation} 
{R_s}/{\gamma^2 c} ={\delta }/{c} \ .
\label{tdelta}
\end{equation}
This corresponds to a single peak within a burst.  Thus it can
be identified with the observed quantity $\delta t$.

If more than one layer overtakes another one then other regions will
collide and other shocks will form. These will be produced in
different parts of the shell that are separated by radial distance of
$\Delta$. Therefore the total observed duration is given by
\begin{equation}
\label{tdur}
t_{dur}={\Delta /c} . 
\end{equation}
We define  the ratio between $\delta$ and $\Delta$
\begin{equation}
\label{deff}
\zeta \equiv {\delta \over \Delta} \le 1.
\end{equation}
Comparison between Eqs. \ref{tdelta}, \ref{tdur} and \ref{deff} shows
that $\zeta$ is also the ratio between the duration of the
individual peaks and the overall duration of the burst.

The external shock which is produced by the ISM sweeps the shell by
the time it arrives at radius
\begin{equation}
R_{\Delta}=l^{3/4}\Delta^{1/4} \ ,
\label{rdelta}
\end{equation}
(see SP).  
Clearly internal
shocks are relevant only if they appear before the external ones. That
is only if
\begin{equation}
\label{isrelevanct}
R_s < R_\Delta \ .
\end{equation}
Using Eqs. \ref{rs} and \ref{rdelta} condition \ref{isrelevanct} can
be translated to:
\begin{equation}
\xi^{3/2} >  \zeta \ .
\label{isrelevance}
\end{equation}
This generalizes the result of SP who found that $\xi>1$ (for
$\zeta=1$) is a necessary condition for internal shocks.  The
condition \ref{isrelevance} can be turned into a condition that
$\gamma$ is sufficiently small:
\begin{equation}
\label{xige1} \gamma \le 280 ~ \zeta^{-1/2}({\frac{t_{dur}}
{10{\rm sec}}})^{-3/8} l_{18}^{3/8}
\end{equation}
We see that internal shocks take place in relatively ``low'' $\gamma$
regime. Fig. 2 depicts the regimes  in the physical parameter space
($\Delta, \gamma$) in which different shocks are possible. It also
depicts an example of a $t_{dur} =$const. line.

\begin{figure}
\psfig{file=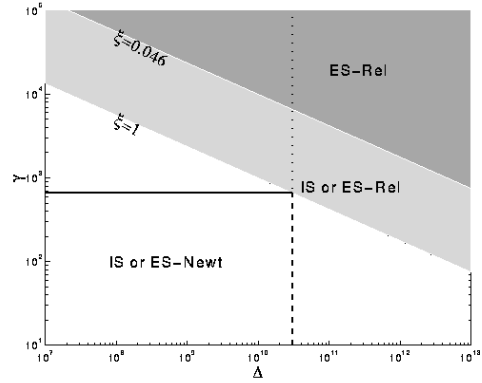,width=220pt}
\caption{Different scenarios in the $\Delta$,(in cm) $\gamma$ plane
for $\zeta\equiv \delta/\Delta=0.01$. Relativistic ES occur for large
$\Delta$ and large $\gamma$ - upper right -above the $\xi=1$ line
(dark gray and light gray regions). Newtonian ES occur below $\xi=1$ -
lower left - white region.  IS occur, if there are sufficient
variation in $\gamma$ below the $\xi=\zeta^{2/3}$ line (light gray and
white regions).  The equal duration $t_{dur}=1$sec curve is shown for
Newtonian ES (solid line) a relativistic ES (dotted line) and IS
(dashed line).  Note that a relativistic ES and an internal shock with
the same parameters have the same overall duration $t_{dur}$ but
different temporal substructure depending on $\delta$.}
\label{fig2}
\end{figure}

Provided that the different parts of the shell have comparable Lorentz
factor differing by factor of $\sim 2$, the shocks are mildly
relativistic. The protons' thermal Lorentz factor will be of order of
unity, and the shocked regions will still move highly relativistically
towards the observer with the approximately the initial Lorentz factor
$\gamma$.  Newtonian shocks have a limiting compression of factor of
$7$ (assuming an adiabatic index of relativistic gas, i.e., $4/3$).  The
density behind the shocks is, therefore, comparable to the density in
front of the shocks.  In-front of the shocks the particle density of
the shell is given by the total number of baryons $E/\gamma m_p c^2$
divided by the co-moving volume of the shell at the radius $R_s$ which
is $4 \pi R_s^2 \Delta \gamma$. The density behind the shock is higher
by a factor of $7$.

The hydrodynamical conditions behind the shocks are therefore:   
\begin{eqnarray}
\label{hydroshocks}
n & = & 7 E/4 \pi \gamma^6 c^4 m_p t_{dur}^3 \zeta^3 \ ,  \\
e & = & n m_p c^2 \ . \nonumber
\end{eqnarray}

\section{Opacity Constrains}

One of the advantages of the internal shock model is that it allows
for short bursts with a lower value of the Lorentz factor $\gamma
$. This eases the baryon purity constraints on source models.
However, too low value of the Lorentz factor or too small emission
radius could lead to a large optical depth. This poses another limitation
on the parameters.

The optical depth for Compton scattering of the photons on the the
shell's electrons at $R_s$ is given by:
\begin{eqnarray}
\tau_e=(E/\gamma) / (4 \pi R_s^2) \sigma_T= 3.9 \times 10^{-5}
\zeta^{-2} \times \\ 
{\left({\gamma \over 100}\right)}^{-5}
{\left({t_{dur} \over 100}\right)}^{-2} E_{51} \ . \nonumber
\end{eqnarray}
The condition  $\tau_e<1$ yields a lower limit on $\gamma$:
\begin{equation}
\label{mingamma}
\gamma \ge 13 {\left({t_{dur}\over 10{~\rm sec}}\right)}^{-2/5} 
\zeta^{-2/5} E_{51}^{1/5} \ ,
\end{equation}
which is not a strong limit.

In addition, the radius of emission should be large enough so that the
optical depth for $\gamma \gamma \rightarrow e^+e^-$ will be less than
unity ($\tau_{\gamma\gamma} < 1$). There are several ways to consider
this constraint.  The strongest constraint is obtained if one demands
that the optical depth of an observed $100$MeV photon will be less
than one (Fenimore, Epstein \& Ho, 1993 ; Woods \& Loeb). Following
these calculations and using Eqs. \ref{rs} to express $R_s$ we find:
\begin{equation}
\label{mingammagamma}
\gamma> 100{\left({\zeta t_{dur} \over 100}\right)}^{-1/4}
\end{equation}

Eqs.  \ref{xige1}, \ref{mingamma} and \ref{mingammagamma} constrains
$\gamma$ to a relatively narrow range. In fact the constraint due to
$\gamma\gamma$ interaction is always more important than the constraint due to
Compton scattering: that is $\tau_{\gamma\gamma} > \tau_e$.  In Fig.
3, we plot the allowed regions in the $\gamma$ and $\delta$ parameter
space.

\begin{figure}
\psfig{file=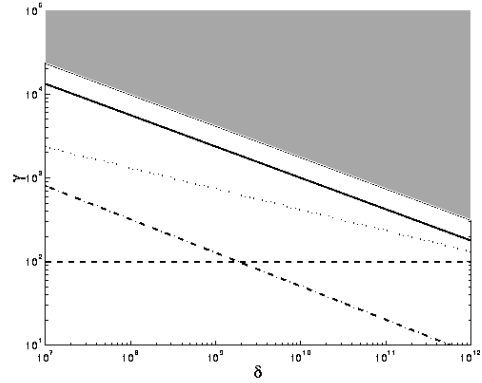,width=220pt}
\caption{Allowed regions for internal shocks in the $\delta$ (in cm), $\gamma$
plane.  Note that the horizontal $\delta$ axis also corresponds to
$\delta t$, the peak duration by a devision by $c$.  Internal shocks
are impossible in the upper right (light gray) region.  The lower
boundary of this region depend on $\zeta \equiv
\delta/\Delta$ are marked by two solid curves, the lower one for $\zeta=1$
and the upper one for $\zeta=0.01$. Also shown are
$\tau_{\gamma\gamma}=1$ for an observed spectrum with no upper bound
(dotted line), $\tau_{\gamma\gamma}=1$ for an observed spectrum with
an upper bound of $100$MeV (dashed line) and  $\tau_e =1$ (dashed-dotted).
The optically thin internal shock region is above the 
$\tau=1$ curves and below the $\xi=\zeta^{2/3}$ (solid) lines. } 
\label{fig3}
\end{figure}

Three main conclusions emerge from the discussion so far. First, if
the spectrum of the observed photons extends beyond 100MeV (as was the
case in the bursts detected by EGRET) and if those high energy photons
are emitted in the same region as the low energy ones then the
condition on pair production, $\tau_{\gamma\gamma}$, is much stronger
than the condition on Compton scattering, $\tau_e$. This increases the
required Lorentz factors. In fact even if the spectrum has an upper
cutoff of $100$MeV then it sets a lower limit of $\gamma\ge 100$.
Second, even the Compton scattering limit (which is independent of the
observed high energy tail of the spectrum) poses a stronger than
expected limit on $\gamma$ for bursts with very short peaks (see
Fig. 3). The Lorentz factor should be as high as $10^3$ to yield a
peak of 0.3msec. Finally, one sees in Fig. 3 that only in a narrow
region in the $(\delta,\gamma)$ plane in which optically thin internal
shocks are produced. The region is quite small if the stronger pair
production limit holds and in particular there is no single value of
$\gamma$ that can produce peaks over the whole range of observed
durations.  The allowed region is larger if we use the weaker limits
on the opacity. But even with this limit there is no single value of
$\gamma$ that produces peaks with all durations. The IS scenario
suggests that bursts with narrow peaks should not have a very high
energy tail and that very short bursts should have  thermal spectrum.

\section{Synchrotron Cooling}

Following SNP we use the physical conditions derived in the previous
section to estimate the cooling of the shocked regions.  In order to
estimate the radiation processes we need the energy density, $e$, the
particle density, $n$, given by equation \ref {hydroshocks} as well as
the magnetic field strength, $B$, and the distribution of electron
Lorentz factors $\gamma _e$. These last two quantities are difficult
to estimate from first principles. Both depend on the microphysics of
particle acceleration and magnetohydrodynamics within the shocks.  Our
approach, following SNP will be to define two parameters, $\epsilon_B$
and $\epsilon _e$ to incorporate these uncertainties. We then constrain
the values of these parameters by requiring the model predictions to
resemble the observed features of GRBs.

The dimensionless parameter $\epsilon_B$ measures the ratio of the
magnetic field energy density to the total thermal energy density, $e$:
\begin{equation}
\label{EpsilonB} 
\epsilon_B\equiv {\frac{U_B}{e}} = {\frac{B^2}{8\pi e}} \ . 
\end{equation}
Using Eqs.  \ref{hydroshocks} and \ref{EpsilonB} we can express $B$ in
terms of $\epsilon_B$ and other observed quantities:
\begin{equation}
B =(720~{\rm G})\epsilon_B^{1/2}\zeta^{-1}
\left({\frac{t_{dur}}{10~{\rm s}}}\right)^{-3/2}
E_{51}^{1/2}({\frac{\gamma }{100}})^{-3}.
\end{equation}

The second parameter $\epsilon_e$ measures the fraction of the total
thermal energy $e$ which is in the form of the electrons' random motions:
\begin{equation}
\epsilon _e\equiv {\frac{U_e}e} \ .
\end{equation}
Since the electrons obtain their random motions via shock-heating, we
make the standard assumption that they develop a power law
distribution of Lorentz factors,
\begin{equation}
N(\gamma _e)\sim \gamma _e^{\bar \beta }\ \ \ {\rm for}\ \gamma _e>\gamma
_{e,min}\;.
\end{equation}

Following SNP we fix the index $\bar\beta \approx -2.5$ and we find 
that the average electron energy density is
\begin{equation}
\label{epsilone} \epsilon_ee=3 \gamma_{e,min}nm_ec^2 \ . 
\end{equation}
Using Eqs. \ref{hydroshocks} and \ref{epsilone} one finds that the
electrons behind the shocks satisfy
\begin{equation}
\label{gammaemin} \gamma_{e,min}= 610\epsilon_e \ .
\end{equation}
The value of $\gamma_{e,min}$ is sufficient for the calculations of the
cooling time scale and efficiency that follow.  The exact details of
the $\gamma_e$ distribution are not needed. 

We can estimate now the Lorentz factor $\hat \gamma_e$, of an electron
that is radiating an energy $h\nu \approx 100$KeV:
\begin{eqnarray}
\label{gamma_hat} 
\hat \gamma _e=\left( \frac{m_ech\nu_{obs}} {\hbar q_e\gamma B}\right)
^{1/2} = 1.1\times10^4\epsilon_B^{-1/4} \zeta^{1/2} \times \\
\left({\frac{ h\nu_{obs}}{100~{\rm KeV}}}\right)^{1/2}
\left({\frac{t_{dur}}{10~{\rm s}}} \right)^{3/4}E_{51}^{-1/4}
\left({\frac{\gamma}{100}}\right)\ . \nonumber
\end{eqnarray}
We first check that electrons with this Lorentz factor are
available. This requires, of course, $\gamma_{min}<\hat \gamma_e$,
which corresponds to
\begin{equation}
\label{maxeer} \epsilon _e<18\epsilon _b^{-1/4} \zeta^{1/2} 
\left({\frac{h\nu_{obs}}{100~%
{\rm KeV}}}\right)^{1/2} \left({\frac{t_{dur}}{10~{\rm s}}}%
\right)^{3/4}E_{51}^{-1/4}
\left({\frac{\gamma}{100}}\right) \ .
\end{equation}
These electrons are always available. A similar situation occurs in a
reverse external shock. But in a forward external shock for short
burst durations (which force high Lorentz factor) the minimal electron
energy is too high and most of the radiation emerges in higher
energies than the observed band.

Using the value of $\hat \gamma _e$ we can estimate the synchrotron
cooling time. The electron's energy divided by the power of
its synchrotron radiation (with the appropriate Lorentz transformation
from the electron's frame to the observer's frame): 
\begin{eqnarray}
t_{syn}=(0.0014 ~{\rm sec})\epsilon _B^{-3/4} \zeta^{3/2} \left(
{\frac{h\nu _{obs}}{100~{\rm KeV}}}\right)^{-1/2} \times \\ 
\left( {\frac \gamma {100}}\right) ^4\left( {\frac{t_{dur}}{10~{\rm
s}}}\right) ^{9/4}E_{51}^{-3/4} \ . \nonumber
\label{tausyn2}
\end{eqnarray}

This cooling time is shorter than the cooling time estimated for
external shocks (SNP) since the magnetic field here is higher. It sets
a lower limit to the variability time scale.  Clearly, the burst
cannot possibly contain spikes that are shorter than its cooling
time. The cooling time, $t_{syn}$ is proportional to $\left( h\nu
\right) ^{-1/2}$ as in the case of ES and in agreement with
observations. This is a general feature of synchrotron radiation.  An
observation of this feature can, therefore, verify that synchrotron
radiation takes place in GRBs but it would not distinguish between the
ES and the IS scenarios.

Using Eqs. \ref{tausyn2} we  calculate the duty-cycle $D$ defined as the
ratio of the cooling time to the total duration: 
\begin{eqnarray} 
\label{dutycyclesyn}
D\equiv {\frac{t_{syn}}{t_{dur}}}=1.4\times
10^{-4}\epsilon _B^{-3/4} \zeta^{3/2} 
\left( {\frac{h\nu _{obs}}{100~{\rm KeV}}}\right)^{-1/2} \times \\
\left( {\frac
\gamma {100}}\right) ^4\left( {\frac{t_{dur}}{10~{\rm s}
}}\right)^{5/4}E_{51}^{-3/4} \ . \nonumber
\end{eqnarray}
The duty cycle increases with burst duration, but it is short enough
to explain the high variability even for bursts as long as $100$sec .
This is different from the ES scenario in which the duty-cycle was
only weakly decreasing with burst duration. Since duration changes by
about four orders of magnitude it might be possible to distinguish
between the two models.  Note, however, that in internal shocks the
synchrotron cooling is so rapid that another process (such as the
acceleration within the shock) could easily be slower and it would
determine the shortest observe time scale. Thus, the strong dependence
of $t_{syn}$ on $t_{dur}$ might be screened and undetectable.

The cooling time scale increases as the magnetic field decreases. 
Most GRBs are highly variable with a duty-cycle of order of $5\%$ or
less.  This sets a lower limit to the value of $\epsilon _B$:
\begin{eqnarray}
\label{lowerlimitsyn}
\epsilon _B\ge 3.8 \times 10^{-4} \zeta^2 \left({\frac{
h\nu_{obs}}{100~{\rm KeV}}}\right)^{-2/3} \times \\
\left({\frac{D}{0.05}} \right)^{-4/3}
\left({\frac{\gamma}{100}}\right)^{16/3} \left({\frac{t_{dur}}{10~{\rm
s}}} \right)^{5/3}E_{51}^{-1} \ . \nonumber
\end{eqnarray}
Unlike ES, the magnetic field here can be quite far from equipartition.
This eases somewhat the requirement  on the processes that build up the
magnetic fields in the shocks and it is possible that the magnetic field
dragged along the fluid from the source might be sufficient.

\section{Inverse Compton Scattering}

Inverse Compton (IC) scattering may modify our analysis in several
ways.  IC can influence the spectrum even if the system is optically
thin (as it must be) to Compton scattering (see e.g. Rybicki  \&
Lightman, 1979).  The effect of IC depends on the Comptonization
parameter $Y=\gamma^2 \tau_e$.  It can be shown (SNP) that $Y$
satisfies:
\begin{eqnarray} 
Y= {\epsilon _e/\epsilon _B}~~~     & \ \ {\rm  if }\ \ & \epsilon_e 
\ll \epsilon _B\\
Y= \sqrt{\epsilon _e/\epsilon_B} & \ \ {\rm  if }\ \ & \epsilon_e
\gg \epsilon _B . \nonumber
\end{eqnarray}
IC is unimportant if $Y<1$ and in this cases it can be ignored.  

If $Y>1$, which corresponds to $\epsilon_e > \epsilon_B$ and to
$Y=\sqrt{\epsilon_e/\epsilon_B}$ then a large fraction of the low
energy synchrotron radiation will be up scattered by IC and a large
fraction of the energy will be emitted via the IC processes. If those
IC up scattered photons will be in the observed energy band then the
observed radiation will be IC and not synchrotron photons. If those IC
photons will be too energetic, then IC will not influence the observed
spectra but as it will take a significant fraction of the energy of
the cooling electrons it will influence the observations in two ways:
it will shorten the cooling time (the emitting electrons will be
cooled by both synchrotron and IC process).  Second, it will influence
the overall energy budget and reduce the efficiency of the production
of the observed radiation. We turn now to each of this cases.

Consider, first, the situation in which $Y>1$ and the IC photons are
in the observed range so that some of the observed radiation may be
due to IC rather than synchrotron emission. This is an interesting
possibility since one might expect that the IC process to ease the
requirement of rather large magnetic fields that is imposed by the
synchrotron process. We show here that, somewhat surprisingly, this is
not the case.

An IC scattering boosts the energy of the photon by a factor
$\gamma^2_e$.  Therefore, the Lorentz factor of electrons radiating
synchrotron photons which are IC scattered on electrons with the same
Lorentz factor and have energy $h \nu$ in the observed range is the
square root of $\hat \gamma_e$ given by Eq. \ref{gamma_hat}.  These
electrons are cooled  both by synchrotron and by IC.  The latter is
more efficient  and the cooling is enhanced by the Compton parameter
$Y$.  We can now estimate the duty-cycle (the ratio of the cooling
time to the overall duration):
\begin{eqnarray}
\label{DIC}
D=0.014 \epsilon_B^{-3/8} \epsilon _e^{-1/2}\zeta^{7/4} {\left( {\frac
\gamma{100}}\right) }^{9/2} \times \\ 
{\left( {\frac{t_{dur}}{10~{\rm sec}}}\right) } ^{13/8}E_{51}^{-7/8} \left( {\frac{h\nu _{obs}}{100~{\rm
KeV}}}\right) ^{-1/4} \ . \nonumber
\end{eqnarray} 
This duty-cycle depends strongly on the duration, a fact that was not
observed. Additionally, it scales with the observed photon's energy as
$(h\nu )^{-1/4}$ while observation are more compatible with $(h\nu
)^{-1/2}$ (Fenimore \etal 1995).  It is unlikely therefore, that the observed
radiation is due to IC process. In addition, inspection of
Eq. \ref{DIC} shows that the magnetic field cannot be very small even
in this case. If $\epsilon_B$ is too small $D$ will become larger than
unity.

Finally, if $Y> 1$ IC will influence the process even if the observed
photons are not produced by IC. It will  speed up the cooling of the
emitting regions and shorten the cooling time, $t_{syn}$ estimated
earlier by  a factor of $Y$.  With this increase in the
cooling rate the duty-cycle limit on the magnetic field become
extremely small.  However, these very low magnetic field values are
impossible since in addition IC also reduces the efficiency by the
same factor, and the efficiency becomes extremely low as described
below.

The efficiency of a burst depends on three factors: first only the
electrons' energy is available. This yields a factor $\epsilon
_e$. Second, there is an additional factor of of $\sqrt{\epsilon
_B/\epsilon _e}$ if the IC radiation is not observed.  Third, only the
energy radiated by electrons with $ \gamma \ge \hat \gamma $ is
observed, and assuming a a power low electron distribution with an
index $\bar \beta =-2.5$ (see SNP and the previous section) this gives
a factor of $(\gamma _{min}/\hat \gamma )^{1/2}$. The total efficiency
is the multiplication of those three factors and assuming that $
\epsilon _B<\epsilon _e$ it is given by:
\begin{equation}
\epsilon _{tot}\ge 0.24\epsilon _B^{5/8}\epsilon _e \zeta^{-1/4}
E_{51}^{1/8}\left( {\frac\gamma {100}}\right) ^{-1/2}
\left( {\frac{t_{dur}}{10~{\rm s}}}\right)^{-3/8} \ .
\end{equation}
The efficiency can be rather high provided that the electrons and the
magnetic field energy density are close to equipartition.  This
efficiency equation sets a stronger lower limit for the magnetic field
than the duty-cycle lower. As in the ES case the efficiency increases
when the burst duration decreases.

\section{Discussion}

We have calculated the hydrodynamic and cooling time scales of GRBs
produced by internal shocks. We have found that internal shocks could
easily explain the observed complicated temporal structure of GRBs: it
directly reflects the temporal behavior of the inner engine that
drives the GRB.  Recall, that this is not the case in the ES scenario
(SP).  The duration is dictated by $\Delta$, the width of the
relativistic shell, which in turn corresponds to the total duration of
the emission of the internal source. The duration of individual peaks
is determined by $\delta$ the length scale over which the conditions
in the shell vary significantly. The length $\delta$, is again
dictated by the temporal scale of variability of the internal source.
This scenario agrees, therefore, with the conclusion of Li \& Fenimore
(1996) who discovered that the mid-peak intervals has a Log-Normal
distribution which indicates a causal relation between peaks and
concluded that the temporal structure is driven by the inner source.

The opacity, and in particular the opacity to pair-production poses a
sever constraint on the parameter space. The strongest constrain
appears if we demand that an observed 100MeV photon can escape freely
from the system (Fenimore, Epstein and Ho, 1993; Woods and Loeb,
1995). The fact that there are such observed photons and that EGRET
observes even higher energy photons suggests that this demand is
reasonable.  If the observed spectrum is unbounded than this yields a
strong limit and only a narrow strip is allowed for optically thin IS
in the $\gamma$, $\delta$ plane. The most worrisome fact is that there
is no single value of $\gamma$ (or $\delta$) which can account for all
kinds of observed bursts and the shape of this allowed region demands
a correlation (in fact anti-correlation) between $\gamma$ and
$\delta$. This is not impossible but it seems unlikely.  Alternatively
this suggests that bursts with short peaks will be optically thick to
$100$MeV photons and such photons won't be observed in those bursts.

If there is an upper cutoff $E_{max}$to the observed spectrum then the
optically thickness problem is weaker and pair-production simply set a
limit of a minimal Lorentz factor of $\sim E_{max}/2 m_e c^2$ which is
$\gamma \approx 100$ for $100$MeV photons. Even in this case the
optical depth to Compton scattering becomes larger than unity for
small values of $\delta$ - which correspond to very short peaks. The
allowed region is wider but once more either there is some correlation
between $\gamma$ and $\delta$ or we should observe optically thick
narrow peak bursts.

Like in the ES scenario the radiation process is most likely
synchrotron.  In spite of the higher energy density in the internal
shock region (higher compared to the densities in the ES scenario) the
magnetic field values required are, close to the equipartition
value. Otherwise, the efficiency will become too low!  Inverse Compton
(IC) is possible - but quite unlikely. Somewhat surprisingly IC
cooling requires magnetic fields that are as high as those required
for synchrotron cooling. Again if these values are  too small then
efficiency constraints will rule out the model.

Synchrotron cooling in the IC scenario suggests a correlation between
the cooling time and the overall duration: $t_{syn} \propto
t_{dur}^{9/4}$. This is a high power and if $t_{syn}$ determines the
duration of the shortest peaks it might be observed since $t_{dur}$
varies by four orders of magnitude.  This should be compared to the
cooling time in the ES scenario which varies like $t_{dur}^{3/4}$.
However, the cooling rate in the IS scenario is so rapid that this
possible correlation may be masked by another process, e.g.  the
acceleration within the shocks, that would be slower and would
determine the duration of the shortest peaks.

\section*{Acknowledgments}
We thank Ramesh Narayan and Jonathan Katz for helpful discussions. This
research was supported by the US-Israel BSF science foundation and by
the Israel NSF.

\end{document}